\documentclass[11pt]{article}

\usepackage[final]{acl}

\usepackage{times}
\usepackage{latexsym}

\usepackage[T1]{fontenc}

\usepackage[utf8]{inputenc}

\usepackage{microtype}

\usepackage{inconsolata}

\usepackage{graphicx}

\graphicspath{{fig/}{./}}

\usepackage{balance} 
\usepackage{enumitem}
\usepackage{multirow}
\usepackage{subcaption}
\usepackage{xcolor}
\usepackage{xspace}
\usepackage{booktabs}
\usepackage{adjustbox}
\usepackage{tabularx}
\usepackage{tcolorbox}
\tcbuselibrary{listings,skins,breakable}
\usepackage{xurl}

\newcommand{\tool}[0]{\textsf{{ABRA}}\xspace}

\newcommand{\mOpus}{Opus 4.8}
\newcommand{\mHaiku}{Haiku 4.5}
\newcommand{\mGPT}{GPT-5.4}
\newcommand{\mMini}{GPT-5.4-mini}

\lstdefinestyle{promptstyle}{
  basicstyle=\scriptsize\ttfamily\linespread{0.8},
  aboveskip=0pt,
  belowskip=0pt,
  breaklines=true,
  columns=fullflexible,
  keepspaces=true,
  extendedchars=true,
  literate=
    {_}{{\_}}1
    {—}{{---}}1
    {–}{{--}}1
    {•}{{\textbullet}}1
    {…}{{\ldots}}1
    {“}{{``}}1
    {”}{{''}}1
    {‘}{{`}}1
    {’}{{'}}1
}

\newtcolorbox{promptbox}[2][]{%
  enhanced, breakable,
  width=\linewidth,
  boxsep=2pt, left=6pt, right=5pt, top=5pt, bottom=5pt,
  fonttitle=\hypersetup{linkcolor=white,urlcolor=white},
  title={#2},
  label={#1},
  lower separated=true,
}

\title{\tool: Agent-Bound UI for Human--Agent Communication beyond Chat with Reusable Views}

\author{Seon Gyeom Kim \\
  Megagon Labs \\
  \texttt{seongyeom@megagon.ai} \\\And
  Hannah Kim \\
  Megagon Labs \\
  \texttt{hannah@megagon.ai} \\}

\makeatletter
\AtBeginDocument{
\let\@oldmaketitle\@maketitle
\renewcommand{\@maketitle}{\@oldmaketitle
  \vspace{-30pt}
  \centering
  \includegraphics[width=\linewidth]{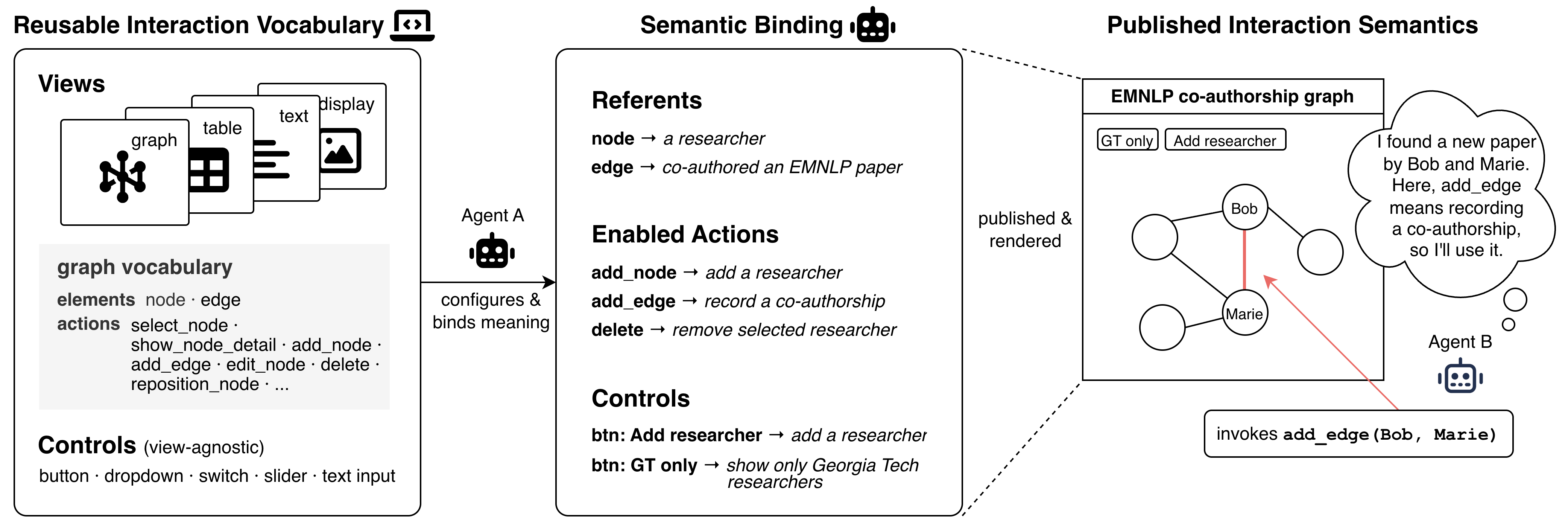}
  \captionof{figure}{
    Agent-Bound UI separates a reusable interaction vocabulary (\S\ref{sec:reusable_vocab}) from task-specific semantic bindings (\S\ref{sec:semantic_binding}). The agent publishes those bindings so that humans and other agents can read and act on them (\S\ref{sec:publish_semantics}).
  }
  \label{fig:agent-bound-ui}
  \vspace{13pt}
 }
}
\makeatother

\begin{document}
\maketitle

\begin{abstract}
Agents can create rich representations, yet these representations are rarely used as a medium for expressing user intent. We present \emph{Agent-Bound UI}, an interaction model that separates reusable interaction structures from task-specific meanings. Agents bind task-specific meanings to interactions in reusable views and publish the resulting bindings as shared state, enabling humans and agents to understand and act on shared surfaces whose meanings vary across tasks. We implement ABRA, a framework-agnostic prototype of Agent-Bound UI, in which users and agents collaborate through chat and shared interactive canvases. We demonstrate ABRA through a representative use case and evaluate four LLM-backed agents on canvas authoring, owner-side consequence resolution, and cross-agent invocation. The code and demo video are available at \url{https://github.com/megagonlabs/abra}.

\end{abstract}

\section{Introduction}
\label{sec:intro}
Agentic AI is increasingly used to assist users with a wide range of tasks~\cite{wang2024survey}, from everyday activities such as drafting emails and planning trips to professional work including cleaning datasets~\cite{chen2026large}, curating knowledge graphs~\cite{10387715}, and writing code~\cite{10.1145/3796507}. Unlike conventional conversational assistants that primarily answer questions, modern agents are connected to external tools and services to actively carry out these tasks, often across multiple coordinating agents. In this setting, chat serves as the primary medium through which humans and agents collaborate. Users specify task goals, review and revise plans, inspect intermediate results, and provide feedback throughout the task lifecycle.

Meanwhile, the way agents present their work to users has steadily evolved. Early agents responded in plain text~\cite{openai2024gpt4technicalreport}, markdown has since become the standard~\cite{10.1145/3696410.3714674}, and more recently interactive HTML-based representations (e.g., artifacts, canvases, and inline widgets) have begun gaining traction~\cite{claude_html}. For example, an agent may present a plan as a flowchart or a set of results as a table rather than as text.

Yet while representations from agents to humans have become increasingly rich, the human-to-agent channel remains largely unchanged, with chat continuing to serve as the dominant interface for user input. This creates a \emph{representation–interaction asymmetry}: the output side is increasingly structured and multimodal, whereas the input side remains constrained to natural language. 

This asymmetry, however, reveals an opportunity: rich representations are already rendered on the user's screen. We therefore ask whether these representations can themselves be made \emph{directly manipulable}~\cite{direct_manipulation}. Rather than introducing a separate input modality, the same surface the agent renders becomes one the user acts on, grounding interaction in the very objects the agent produced. 
The agent can then interpret these interactions as expressions of user intent grounded in task representations, incorporating them into its reasoning.

The effectiveness of chat and direct manipulation depends on the \emph{verbalizability of user intent}. When intent can be readily expressed in language, chat is concise and effective. However, when intent is difficult to verbalize or closely tied to a rendered representation, text-only interaction can become inefficient or ambiguous~\cite{iai}.

Recent approaches that make agent-rendered representations interactive can be broadly grouped into three categories. Generative approaches~\cite{chen-etal-2026-generative} synthesize interfaces for individual tasks, enabling long-tail scenarios that require novel interaction surfaces, but limiting reuse across tasks. Composable ones~\cite{a2ui_blog} assemble interfaces from reusable widgets and controls, making them effective for routine interactions, but providing limited support for rich, task-specific manipulation over representations. Task-specific systems~\cite{kim-etal-2025-aipom,shao2026collaborative,cocoa2026chi} provide richer interaction surfaces tailored to particular workflows, but their interactions remain coupled to the targeted application.

Taken together, these systems leave an important regime underserved: tasks that require direct manipulation over recurring representations, where the surface itself is reusable but the meaning of each manipulation changes across tasks. For example, the same graph view may represent a project plan in one task and a social network in another; connecting two nodes creates a dependency between steps in the first and represents a social relationship in the second.
Supporting this regime requires separating reusable interaction structures from task-specific semantics and binding those semantics to each task. Once meanings vary across uses, the binding itself can no longer remain implicit: it must be published as shared, observable state. 
This allows any participant, human or agent, to inspect the available interactions and understand their meanings when interpreting or performing manipulations.
In particular, as tasks are increasingly shared among agents, a second agent can use an interaction surface created by another agent because its interaction semantics are explicitly shared.

To this end, we present \emph{Agent-Bound UI}, an interaction model that turns reusable interactive views into task-specific communication surfaces by allowing agents to bind and publish interaction semantics. We realize this model in \tool (\textbf{A}gent-\textbf{B}ound \textbf{R}eadable inter\textbf{A}ctions) prototype. Our contributions are as follows:
\begin{itemize}[noitemsep,topsep=2pt]
\item \textbf{Agent-Bound UI}, an interaction model that enables agents to bind and publish interaction semantics over reusable interactive views.
\item \textbf{\tool}, an open, framework-agnostic prototype in which multiple agents and users collaborate over shared interactive canvases.
\item A case study and experiments demonstrating agent-bound interaction in practice.
\end{itemize}

\section{Related Works}
\label{sec:relwork}
\subsection{Generative and Composable Interfaces}
\label{sec:gen-ui}

Generative approaches\footnote{We use the term \emph{generative} to refer to approaches that generate the interface as a medium for human-agent communication; we exclude systems in which the generated interface or application is itself the task deliverable (e.g., prompt-to-app builders).}
 synthesize the interface itself, typically turning a chat request into a generated interactive artifact: a page produced in place of a response~\cite{GenerativeGUI,chen-etal-2026-generative,leviathan2026generative}, or an application that persists and evolves with the task~\cite{duet-ui,xie2026softwarecontentdynamicapplications}. Either way, what a manipulation means is frozen in the generated code, and changing it means regenerating.
\citet{jelly} make the artifact malleable through an evolving data model, but each manipulation still means a fixed, system-defined edit.

Composable approaches provide reusable UI primitives that agents assemble into task-specific interfaces. Systems such as A2UI~\cite{a2ui_blog} assemble task-specific interfaces from component catalogs, enabling reuse of interface components rather than interaction semantics. Adaptation, however, still occurs primarily through interface composition, with richer representation-level interactions often requiring additional task-specific components or application logic. In contrast, Agent-Bound UI keeps the interaction structure reusable while moving adaptation to task-specific semantic bindings.

\subsection{Task-Specific Interaction Surfaces}
\label{sec:task-specific-surfaces}

While generative and composable approaches allow agents to create or assemble interfaces, task-specific systems provide developer-authored interaction surfaces tailored to particular workflows~\cite{shao2026collaborative}.
Examples include writing~\cite{ABScribe,Beyondchat,chi2025textoshop}, agentic coding IDEs, data analysis and visualization~\cite{10292609,waitgpt,11300819}, planning~\cite{cai-etal-2024-low-code,kim-etal-2025-aipom,cocoa2026chi,he2026ambipom}, and sensemaking~\cite{graphologue,sensecape}.
These systems support rich manipulation, but both the interaction surface and the meaning of its interactions are fixed for the target application.

Unlike these dedicated systems, MCP Apps~\cite{mcp_apps} bring interactive surfaces into chat environments through developer-authored components. However, interaction semantics remain defined by the underlying implementation and encapsulated within tool-agent interactions, rather than published as shared state. Agent-Bound UI instead binds and publishes interaction meanings at task runtime, allowing other participants to inspect and use the same interaction surface.

\subsection{Agent–UI Communication Protocols}
Several protocols standardize the wiring around an agent: \citet{agui} (from CopilotKit) carries state and events between an agent and a frontend, A2A~\cite{a2a} carries messages between agents, and MCP connects agents to tools, with MCP Apps as its interface extension. These transports move state, messages, and events, but say nothing about what an interaction means. 
Agent-Bound UI defines that missing layer independently of the underlying transport. Because bindings and events are represented as ordinary messages and state, a canvas can be implemented over any of these protocols, subject to their respective capabilities.

\section{Agent-Bound UI}
\label{sec:interaction_model}
Agent-Bound UI models an interaction as a reusable interaction structure paired with a task-specific semantic binding (Fig.~\ref{fig:agent-bound-ui}). We define a \emph{view} as a reusable interaction vocabulary over a particular representation, while a \emph{canvas} instantiates one or more views with task-specific content and interaction semantics.

We distinguish user gestures from agent-facing actions: a gesture describes how a user performs an interaction (e.g., click or drag) on an element (e.g., a node or column), while an action captures the operation communicated to the agent (e.g., edit a cell or add an edge). The meaning of an action, however, depends on the task context. Agent-Bound UI separates these layers by allowing views to define reusable interaction vocabularies while agents bind task-specific meanings to these actions and publish the resulting bindings as shared canvas state.

\subsection{A Reusable Action Vocabulary}
\label{sec:reusable_vocab}

Agent-Bound UI builds on the observation that many interaction structures recur across tasks and treats them as a reusable vocabulary rather than a task-specific implementation. Each view defines a set of available elements and actions over those elements, providing a stable interaction substrate~\cite{interaction_substrate} that can be reused across tasks. Because these actions describe operations rather than task meanings, the same vocabulary can support different tasks through semantic binding.

The vocabulary may also include a bounded set of composable controls, such as buttons and dropdowns. These controls provide additional interaction elements that can acquire task-specific meanings through the same binding mechanism.

\subsection{Semantic Binding and Action Handling}
\label{sec:semantic_binding}

The interaction vocabulary is task-agnostic. To specialize it for a particular task, the agent binds view elements to task objects (referents), selects the applicable actions, and assigns task-specific meanings to those actions.
This binding can apply both to interactions over existing representations (e.g., editing a table cell or connecting graph elements) and to composable controls (e.g., a button triggering a task-specific operation).

Semantic binding is independent of how an action is resolved. In Agent-Bound UI, actions follow one of two handling modes. A \textbf{local effect} is handled directly by the view when the interaction can be completed from view state alone, such as changing a selection or updating presentation state. In contrast, a \textbf{consequence} delegates resolution to the agent when completing the action requires task-specific reasoning. 
This separation keeps the interaction vocabulary reusable while allowing task-specific behaviors to emerge through semantic binding and agent resolution.

\begin{figure*}[t]
\centering
 \includegraphics[width=.95\linewidth]{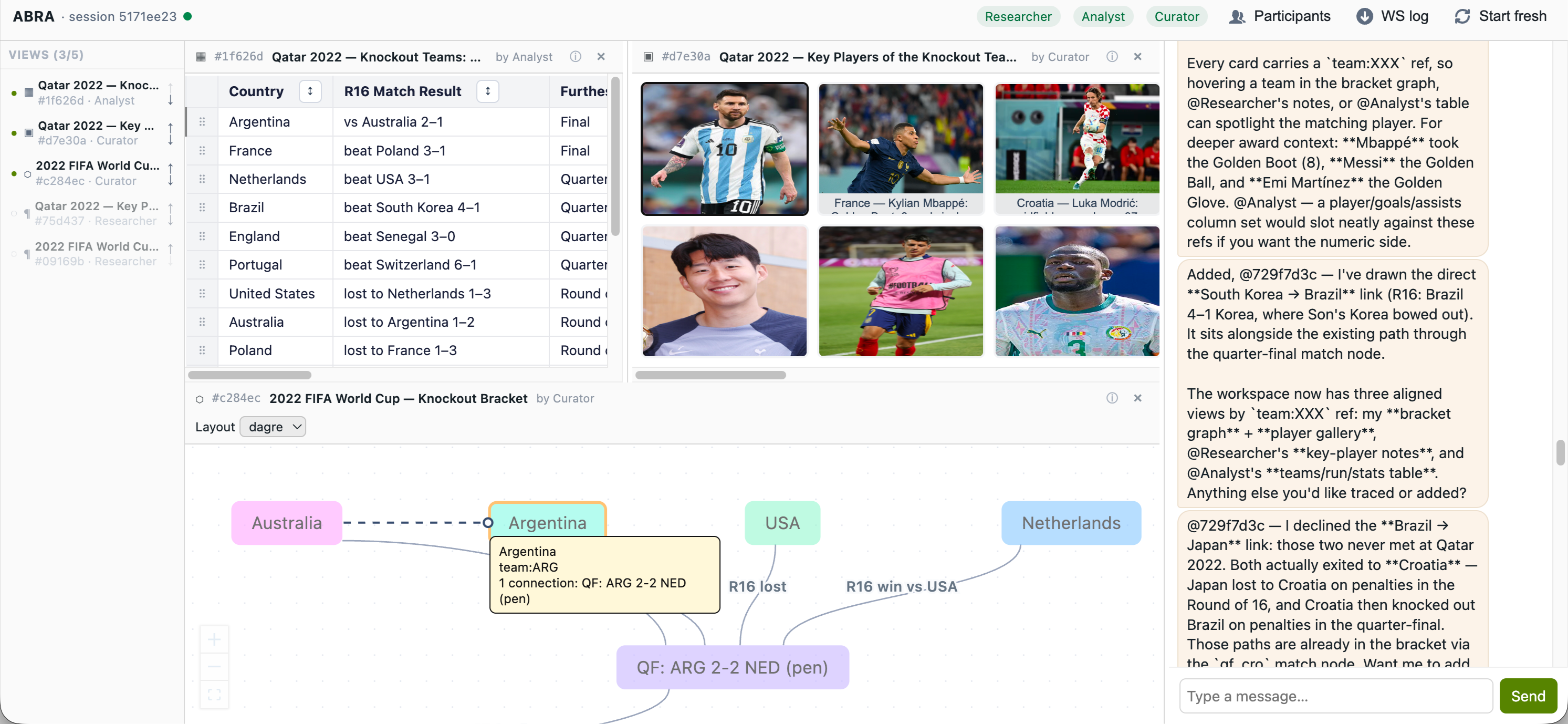}
 \caption{The \tool prototype: a user and multiple agents collaborate
not only through chat but also on shared interactive canvases that
the agents create and co-curate. Hovering an element highlights
co-referent elements across the different views, and consequences are
resolved by the agent that owns each canvas.}
 \label{fig:abra-system}
\end{figure*}

\subsection{Published Interaction Semantics}
\label{sec:publish_semantics}

Because interaction meanings are task-specific, Agent-Bound UI publishes semantic bindings as part of the shared canvas state rather than leaving them as private agent logic. A canvas state includes the bindings between view elements and task referents, which actions are enabled for the canvas, and the task-specific meanings associated with actions. Making these bindings observable allows not only the agent that created the canvas, but also humans and other agents to inspect the available interactions and understand their intended meanings.

Publishing thus turns a single-agent convenience into a multi-agent necessity: one agent can create a canvas and publish task-specific interaction bindings, and a second agent can later interpret or invoke those actions based on their published meanings, with no private channel between them.

\section{\tool System}
\label{sec:system}
\subsection{Realization of Agent-Bound UI}
\label{sec:system_overview}

\tool serves as a reference prototype of the Agent-Bound UI interaction model. It realizes the model through three components: (1) view definitions, (2) canvas specifications, and (3) an agent interface.
Views provide reusable interaction structures, while canvas specifications instantiate these structures with task-specific content and interaction bindings provided by agents.
These bindings are published as shared canvas state, which agents can access and manipulate through tool calls.

The prototype provides a unified interaction environment with a chat panel for natural language interaction and a canvas panel for direct manipulation interaction on agent-rendered views (Fig.~\ref{fig:abra-system}).

\begin{table*}[!t]
 \centering
  \small
  \setlength{\tabcolsep}{5pt}
  \renewcommand{\arraystretch}{1.15}
  \begin{tabularx}{\textwidth}{@{}l l X X@{}}
    \toprule
    \textbf{View} & \textbf{Elements} & \textbf{Actions with Local Effects} (view-executed) & \textbf{Actions with Consequences} (agent-resolved)\\
    \midrule
    \texttt{graph} & node, edge
      & highlight; select; show detail; reposition / pan / zoom / fit; layout
      & add node; add edge; edit node; delete selection\\
    \addlinespace[2pt]
    \texttt{table} & row, column, cell
      & highlight; select; block-select; reorder; sort; filter
      & add row; add column; edit cell; edit column; delete selection; clear\\
    \addlinespace[2pt]
    \texttt{text} & paragraph, line
      & highlight; select; collapse / expand; find
      & add line; edit line; delete selection\\
    \addlinespace[2pt]
    \texttt{display} & item
      & highlight; focus; change layout; fit
      & \textit{none} (read-only)\\
    \bottomrule
  \end{tabularx}
  \caption{The four supported views and their interaction vocabularies in \tool. Views define addressable elements and actions, while agents bind task-specific meanings to enabled actions. Actions are handled either locally by views or through agent-resolved consequences.}
  \label{tab:views}
  \vspace{-1em}
\end{table*}

\subsection{View and Canvas Specification}
\label{sec:view_canvas_spec}

\tool represents reusable interaction vocabularies through view definitions. Each view definition specifies the element schemas and actions supported by the view, along with view-specific metadata. In the current prototype, each action is associated with a predefined execution mode: either a local effect handled directly by the view or a consequence resolved by the agent, for simplicity. Gesture mappings are implemented within each view and remain independent of task-specific semantic bindings. The prototype currently supports four view types: graph, table, text, and display views, summarized in Table~\ref{tab:views}.

A canvas specification defines a task-specific canvas instance. Agents provide these specifications when creating canvases, including selected views, task content, element-referent bindings, enabled actions, task-specific semantic bindings, and optional controls or presentation settings.

\subsection{Agent Tool Interface}
\label{sec:canvas_tools}

The prototype implements the agent-canvas interface through structured tool calls. This choice provides a uniform mechanism for agents to create, inspect, and operate on shared canvas states without relying on a specific agent platform.

A typical canvas lifecycle begins with \texttt{create\_canvas}, where an agent provides a canvas specification to initialize a shared canvas instance it owns. Agents can inspect the current state and available actions through \texttt{read\_canvas}. Agents can modify canvas either directly via \texttt{update\_canvas} or by invoking enabled actions via \texttt{fire\_action}. For interactions that require agent resolution, the resulting consequences are applied through canvas updates. In the current prototype, agent resolution is handled by the owner agent. The resulting state is shared with all participants. When the task concludes, the owner retires the canvas with \texttt{close\_canvas}.

\subsection{Prototype Architecture}
\label{sec:architecture}

The prototype is implemented with React-based views, a Node.js canvas hub, and agent adapters in Python. The canvas hub maintains canvas states and synchronizes updates among participants, while views and adapters provide client-specific access to the shared environment.\footnote{This architecture provides one realization of the shared canvas model. The interaction model does not depend on this deployment choice; the canvas representation and agent interface can be realized in other agent environments.}

For demonstration, we integrate multiple agents implemented with different agent SDKs. These agents can inspect published canvas states and interact with canvases through the provided tools.

\section{Use Case}
\label{sec:use_case}
We demonstrate ABRA through a collaborative World Cup analysis task involving a user and three specialized agents (Fig.~\ref{fig:abra-system}). Rather than producing a static report, the agents progressively construct a shared workspace comprising a knockout-stage table, tournament bracket, player gallery, and explanatory notes. Analyst maintains the tournament table, Curator the bracket and gallery, and Researcher the explanatory notes. The canvases remain active coordination surfaces where both users and agents communicate by manipulating shared representations rather than solely through chat.

The interaction begins with the agents proposing an initial organization of the workspace, which the user refines directly on the rendered representations. For example, after the user abbreviates Forward to FW in a table cell, the agent proposes applying the same convention throughout the column, interpreting the edit as a refinement rather than a literal cell replacement. Likewise, the user connects nodes in the tournament graph to express relationships directly instead of describing them. Not every manipulation, however, is accepted: though the graph view supports edge creation, the owner agent rejects a connection that violates the tournament structure. Thus, views define what interactions users can attempt, while agents determine which are meaningful for the current task.

The shared canvases also serve as the medium for inter-agent coordination. When Curator requires player statistics, it asks Analyst to extend the existing table instead of modifying it directly or duplicating the information. 
Once the updated table is published, Curator then reuses the information in its own canvases. 
Shared referent bindings keep teams and players synchronized across the table, graph, gallery, and notes, allowing participants to move between representations while maintaining a consistent interpretation of each entity.

\section{Experiments}
\label{sec:experiments}
Agent-Bound UI assumes that agents can use our interaction vocabulary to author, interpret, and manipulate task-specific canvases. In particular, agents should be able to (1) author canvases, (2) resolve consequences on canvases they own, and (3) interpret and invoke interaction semantics published by other agents. 
We evaluate these capabilities in isolation rather than evaluating end-to-end task performance. This allows us to assess whether an agent can correctly ground and execute the required operations independently of longer interaction dynamics.

\subsection{Tasks}
\label{sec:exp-tasks}

\paragraph{Canvas authoring}
The agent receives a natural language request and creates a task-specific canvas. We construct 45 cases covering 15 canvas configurations, with four single-view and 11 multi-view configurations, each with three domain variations. 
Each case additionally includes an interactive variant that requires direct manipulation.

\paragraph{Consequence resolution} The agent receives a pre-populated canvas that it owns with a pending consequence and must resolve it through \texttt{update\_canvas}. We construct 33 cases covering 11 consequence-producing actions across three domain variations.

\paragraph{Cross-agent invocation} The agent receives a user request over a canvas owned by another agent and must invoke the appropriate published interaction through \texttt{fire\_action}. We construct 33 cases covering 11 actions across three domain variations.

\subsection{Experimental Setup}
\label{sec:exp-setup}

Each case is run three times per model, with a fresh agent instance for every trial. The agent receives a task-specific context and a single input, and is allowed one response turn to submit canvas operations. The context differs by task: canvas authoring provides a user request; consequence resolution provides a canvas state owned by the agent with a pending consequence; and cross-agent invocation provides a canvas state owned by another agent with published interaction semantics together with a user request. This one-turn setting evaluates whether agents can correctly ground and execute canvas operations from the available context without relying on iterative interaction.

We compare four LLM-backed agents using models from two families: Claude \mOpus ~and \mHaiku, and \mGPT~ and \mMini.
Agent performance is evaluated through automatic structural checks and an LLM-based evaluator using predefined criteria. Details are provided in \S\ref{app:eval-detail}.

\begin{table}[t]
\centering\small
\setlength{\tabcolsep}{3.5pt}
\begin{adjustbox}{max width=\columnwidth}
\begin{tabular}{@{}lcccc@{}}
\toprule
 & \mOpus & \mHaiku & \mGPT & \mMini \\
\midrule
\multicolumn{5}{@{}l}{\emph{\texttt{\textbf{create\_canvas}}} (\(n = 135\))} \\
Pass rate (\%)            & 91.9 & 93.3 & 97.8 & 88.9 \\
\quad Missing operation           & 2    & 4    & 3    & 13   \\
\quad Schema/shape        & 9    & 5    & 0    & 2    \\
\midrule
\multicolumn{5}{@{}l}{\emph{\texttt{\textbf{update\_canvas}}} (\(n = 99\))} \\
Pass rate (\%)            & 98.0 & 100.0 & 100.0 & 98.0 \\
\quad Content corruption  & 2    & 0     & 0     & 2    \\
\midrule
\multicolumn{5}{@{}l}{\emph{\texttt{\textbf{fire\_action}}} (\(n = 99\))} \\
Pass rate (\%)            & 100.0 & 91.9 & 97.0 & 66.7 \\
\quad Missing operation           & 0     & 8    & 3    & 1    \\
\quad Incorrect operation          & 0     & 0    & 0    & 24   \\
\quad Schema/shape        & 0     & 0    & 0    & 8    \\
\bottomrule
\end{tabular}
\end{adjustbox}
\caption{Structural pass rates and failure counts. Each model was evaluated on 3 runs per case; pass rate is the percentage of runs completed without structural errors, and the remaining rows report failure counts.
\emph{Missing operation}: did not issue the required canvas operation.
\emph{Incorrect operation}: issued a different canvas operation than required.
\emph{Schema/shape}: the canvas spec or action parameters were malformed. 
\emph{Content corruption}: the update altered content beyond the expected consequence.
}
\label{tab:tier1}
\end{table}

\begin{table}[t]
\centering\small
\setlength{\tabcolsep}{3.5pt}
\begin{adjustbox}{max width=\columnwidth}
\begin{tabular}{@{}lcccc@{}}
\toprule
 & \mOpus & \mHaiku & \mGPT & \mMini \\
\midrule
\multicolumn{5}{@{}l}{\emph{\texttt{\textbf{create\_canvas}}}} \\
view selection    & 50 & 53 & 48 & 33 \\
view recall       & 77 & 78 & 72 & 56 \\
action selection  & 80 & 70 & 47 & 39 \\
action recall     & 90 & 81 & 57 & 53 \\
\addlinespace
referent coverage & 99 & 99 & 96 & 92 \\
element binding   & 100 & 100 & 100 & 100 \\
spec format       & 99 & 94 & 65 & 92 \\
element format    & 100 & 100 & 98 & 100 \\
value             & 97 & 93 & 93 & 87 \\
action meaning    & 88 & 76 & 90 & 92 \\
title            & 100 & 100 & 100 & 100 \\
ref coordination & --  & 56 & 100 & 93 \\
\addlinespace[8pt]
\multicolumn{5}{@{}l}{\emph{\texttt{\textbf{fire\_action}}}} \\
action selection  & 100 & 100 & 100 & 100 \\
target reference  & 100 & 100 & 100 & 97 \\
spec format       & 94 & 93 & 93 & 85 \\
element format    & 83 & 83 & 83 & 98 \\
value             & 95 & 97 & 100 & 98 \\
style convention  & 96 & 93 & 95 & 88 \\
\bottomrule
\end{tabular}
\end{adjustbox}
\vspace{-4px}
\caption{Per-criterion scores (\%). Exact selection measures whether the full expected set was produced, whereas recall measures the fraction of expected views or actions included. The rest are LLM-judge scores over applicable runs.}
\label{tab:criteria}
\end{table}

\subsection{Results}

Table~\ref{tab:tier1} reports structural success rates across canvas tasks, measuring whether each model produced valid canvas operations.
For \texttt{create\_canvas}, pass rates ranged from 88.9\% to 97.8\%, with the remaining failures attributable to schema or shape errors.
For \texttt{update\_canvas}, pass rates ranged from 98\% to 100\%.
Its four failures, two from Opus and two from GPT-5.4-mini, modified content beyond the expected outcome.
Performance varied more substantially for \texttt{fire\_action}.
Opus, GPT-5.4, and Haiku passed in 91.9\% to 100\% of runs, whereas GPT-5.4-mini passed in 66.7\%.
Of its 33 failures, 24 called \texttt{update\_canvas} instead of \texttt{fire\_action}, eight omitted the required \texttt{path}, and one produced no action.

Table~\ref{tab:criteria} characterizes the generated canvases in greater detail (c.f., \S\ref{app:judge-criteria}).
Models selected appropriate views but often omitted parts of the expected set.
View recall ranged from 56\% to 78\%, while exact all-or-nothing selection ranged from 33\% to 53\%. 
This gap indicates that multi-view requests often produced some, but not all, expected views.
GPT-5.4-mini had the lowest selection score at 33\%.

Element binding reached 100\% for every model, but other judged criteria showed larger differences. 
Spec format ranged from 92\% to 99\% for three models but fell to 65\% for GPT-5.4, which frequently placed the \texttt{config} block inside \texttt{content} rather than at the root. 
Although allowed by the contract, this structure is anomalous. Action meaning fell to 76\% for Haiku, which often restated the action name rather than describing its task-specific effect.
Among the optional criteria, informative titles were nearly universal, whereas cross-view \texttt{ref} coordination appeared in only \(0/9/12/15\) runs.
Opus did not coordinate referents across views in any run.

Among structurally valid \texttt{fire\_action} runs, judged performance was near ceiling for all models.
Agents that successfully invoked an action generally selected the correct published binding and grounded the correct element. 

These results show that our canvas contract enables the evaluated LLM agents to reliably produce and manipulate executable interfaces, while revealing remaining challenges in open-ended specification following and semantic grounding.

\section{Conclusion}
\label{sec:conclusion}
We presented \tool, a prototype of the Agent-Bound UI interaction model for shared interactive canvases. Our demonstration and experiments show that evaluated LLM agents can effectively operate over published interaction semantics, providing a foundation for richer human–agent interaction and cross-agent collaboration.

\section*{Limitations}
Agent-Bound UI introduces a mechanism for binding and publishing interaction semantics, but several questions remain open. First, determining reusable interaction vocabularies that provide broad task coverage remains challenging; future work will investigate methods for discovering common views and actions from existing interactive systems and evaluating their coverage over diverse task benchmarks. Second, while published semantics are intended to make interactions understandable to both human and agent participants, we have not yet evaluated how effectively users interpret these semantics or how they affect the overall interaction experience. Future user studies will examine how interaction semantics should be presented, and whether agent-bound interactions improve user experience and task outcomes compared with chat-based interaction.



\bibliography{custom,anthology-1,anthology-2}

\appendix
\lstdefinestyle{judgeprompt}{%
  basicstyle=\ttfamily\footnotesize,
  breaklines=true, breakindent=0pt, columns=fullflexible,
  keepspaces=true, showstringspaces=false,
  frame=single, framesep=4pt, xleftmargin=2pt, xrightmargin=2pt
}


\begin{table*}[t]
\centering\small
\begin{tabularx}{\linewidth}{@{}l >{\raggedright\arraybackslash}X l@{}}
\toprule
\textbf{Criterion} & \textbf{Question} & \textbf{Scale} \\
\midrule
\multicolumn{3}{@{}l}{\emph{\texttt{\textbf{create\_canvas}}}} \\
\texttt{view\_selection} & Did the run build the view(s) the case expects? & recall 0--1, exact \\
\texttt{referent\_coverage} & Are the task's information-unit kinds present at all (placement aside)? & covered/total \\
\texttt{element\_binding} & Does each element kind stand for the referent kind the task intends? & pass/na \\
\texttt{spec\_format} & Is the spec structurally whole? (the home for any structural anomaly) & pass \\
\texttt{element\_format} & Are ids and paths written in the contract's form? & pass/na \\
\texttt{value} & Does the content deliver the right entities and relations, complete, nothing invented? & 0--1, defects \\
\texttt{action\_selection}\textsuperscript{$\ddagger$} & Are the requested manipulations enabled in config.actions? & recall 0--1 \\
\texttt{action\_meaning}\textsuperscript{$\ddagger$} & Does each published meaning say what its action accomplishes for this task? & good/total \\
\addlinespace
\texttt{title}\textsuperscript{$\dagger$} & Is the title informative for the task? & pass/na \\
\texttt{ref\_coordination}\textsuperscript{$\dagger$} & Do elements shared across views carry the same ref so the views link? & pass/na \\
\addlinespace[7pt]
\multicolumn{3}{@{}l}{\emph{\texttt{\textbf{fire\_action}}}} \\
\texttt{action\_selection} & Is the fired action the published one whose meaning matches the problem? & pass, defects \\
\texttt{target\_reference} & Does the path denote the element the problem is about? & pass/na, defects \\
\texttt{spec\_format} & Does the call carry the fields the action needs? & pass \\
\texttt{element\_format} & Are the IDs and paths the call names (its path and any value ref) in the contract's form? & pass/na \\
\texttt{value} & Does the authored value solve the problem? & pass/na \\
\texttt{style\_convention} & Does authored content follow the canvas's established formatting? & pass/na \\
\bottomrule
\end{tabularx}
\caption{Judge rubrics for \texttt{create\_canvas} (top) and \texttt{fire\_action} (bottom). $\dagger$ denotes optional criteria reported separately from failure criteria; $\ddagger$ applies only to the \texttt{+action} variant of \texttt{create\_canvas}.
}
\label{tab:judge-rubric}
\end{table*}


\section{Evaluation Details}
\label{app:eval-detail}

We combine deterministic checks with an LLM-based evaluation (from a separate model family, Gemini 3.1 Pro) to assess agent outputs. 
Deterministic checks verify contract-level constraints, including schema validity and 
required output formats, as well as task-specific expected outcomes. For 
\texttt{create\_canvas}, we additionally check whether the generated view and action 
types match the expected specification. For \texttt{update\_canvas}, we verify that 
the revised canvas matches the expected content after applying the update. For 
\texttt{fire\_action}, we check whether the fired action type matches the expected 
action.

The LLM evaluator handles open-ended criteria that cannot be reliably determined 
through deterministic checks, including semantic grounding, content correctness, 
and style conventions.

\subsection{Judge Prompts}
\label{app:judge-prompts}


\begin{lstlisting}[style=judgeprompt,title={System Prompt}]
You evaluate what an agent wrote to a shared interactive canvas.
{what a canvas is: the view fixes the interaction structure; the agent authors, per task, what each
 interaction MEANS and publishes it as readable canvas state, so other agents and humans can use a
 canvas they did not create. You grade the authored part against the fixed part.}

FIXED BY THE VIEW, identical across all tasks:
{for each of the four views (graph, table, text, display): its element types, its content shape, and a
 closed catalogue of actions over them; * marks a consequence action that changes content, the rest are
 coordination actions that only change what is shown}

AUTHORED BY THE AGENT, per task:
{which view(s) to instantiate; a title; the content, elements keyed by id, with an optional per-element
 ref that links equal referents across views; the published action meanings; optionally, composable
 controls and presentation settings}

OWNERSHIP:
{one owner per canvas (the agent that created it) alone resolves the consequences others fire; in
 exclusive mode a non-owner only fires published actions, in collaborative mode it may co-author content
 under a lock but never changes the published meanings}

Grade only the authored part against the fixed part. Reply with one compact JSON object in the shape the
request specifies, nothing else.
\end{lstlisting}


\begin{lstlisting}[style=judgeprompt,title={Prompt for Canvas Authoring}]
Grade a create_canvas run: an agent, given a request, instantiated one or more views and filled them for the task.

REQUEST:
{request}

EXPECTED VIEWS (for view_selection --- the case expects these; an extra appropriate view is fine):
{views}

{action_block}THE CANVAS(ES) THE AGENT BUILT (view, title, content with any per-element refs, config):
{authored}

{rules}

CRITERIA --- answer each in the JSON shape shown after it:
{criteria}

Return ONE JSON object, nothing else, no markdown fences:
{output}
\end{lstlisting}


\begin{lstlisting}[style=judgeprompt,title={Prompt for Cross-Agent Invocation}]
Grade a fire_action run: an agent that does not own a canvas fired ONE of its published actions to solve a problem, for the owner to carry out.

THE PROBLEM:
{problem}

THE CANVAS --- its content, and the actions it publishes (id -> the meaning its owner bound; the agent chooses from these):
{canvas}

THE FIRE THE AGENT COMPOSED:
  action: {action}
  path:   {path}
  value:  {value}
WHAT IT ADDRESSES: {where}

{rules}

CRITERIA --- answer each in the JSON shape shown after it:
{criteria}

Return ONE JSON object, nothing else, no markdown fences:
{output}
\end{lstlisting}


\subsection{Evaluation Criteria}
\label{app:judge-criteria}

Table~\ref{tab:judge-rubric} summarizes the criteria used for canvas authoring (\texttt{create\_canvas}) and cross-agent invocation (\texttt{fire\_action}).

\end{document}